\documentclass[10pt,letterpaper]{article}
\usepackage{opex3}
\usepackage[centerlast]{subfigure}

\begin{document}

%%%%%%%%%%%%%%%%%% title page information %%%%%%%%%%%%%%%%%%
\title{Sub-kHz-level relative stabilization of an intracavity doubled continuous wave optical parametric oscillator using Pound-Drever-Hall scheme}

\author{O. Mhibik, D. Pab\oe{}uf, C. Drag, and F. Bretenaker}

\address{Laboratoire Aim\'e Cotton, CNRS-Universit\'e Paris Sud 11, 91405 Orsay Cedex, France}

\email{Fabien.bretenaker@lac.u-psud.fr} %% email address is required



\begin{abstract}
We report the relative frequency stabilization of an intracavity frequency doubled singly resonant optical parametric oscillator on a Fabry-Perot \'etalon. The red/orange radiation produced by the frequency doubling of the intracavity resonant idler is stabilized using the Pound-Drever-Hall locking technique. The relative frequency noise of this orange light, when integrated from 1~Hz to 50~kHz, corresponds to a standard deviation of 700~Hz. The frequency noise of the pump laser is shown experimentally to be transferred to the non resonant signal beam. 
\end{abstract}

 % REPLACE WITH CORRECT OCIS CODES FOR YOUR ARTICLE

%%%%%%%%%%%%%%%%%%%%%%% References %%%%%%%%%%%%%%%%%%%%%%%%%

%%%%%%%%%%%%%%%%%%%%%%%%%%  body  %%%%%%%%%%%%%%%%%%%%%%%%%%
\section{Introduction}
The recent demonstration of solid-state quantum memories \cite{Riedmatten2008,Hedges2010,Clausen2011,Saglamyurek2011} has raised the possibility to implement solid-state quantum repeaters in quantum key distribution systems. This is an important step towards the increase of the distance over which quantum communications can be achieved. Such solid-state quantum memories are based on the use of the long-lived quantum coherences which exist in rare earth ions cooled down at cryogenic temperatures. However, among the possible rare earths, those which exhibit the longest hyperfine coherence lifetimes (i.~e., the longest possible storage times) and the longest optical coherence lifetimes (i.\ e., the longest available durations for pulse sequences aiming at their coherent manipulation) are europium and praseodymium. For example, Eu$^{3+}$ has been shown to exhibit a dephasing time of 2.6\ ms for its optical transition at 580\ nm, leading to optical homogeneous linewidths as narrow as 122\ Hz \cite{Equall1994}. Similarly, Pr$^{3+}$ has been shown to exhibit homogeneous linewidths of the order of 1\ kHz for its optical transition at 606\ nm \cite{Equall1995}. These long-lived coherences also constitute an interesting resource for applications of rare earth ions to quantum computing \cite{Longdell2004-1,Longdell2004-2,Rippe2005}. However, up to now, all these demonstrations based on Pr$^{3+}$ and Eu$^{3+}$ ions use frequency stabilized dye lasers. Indeed, the coherent manipulation of optical coherences exhibiting lifetimes in the ms range relies on the use of a laser source with a linewidth in the kHz range over durations of the order or longer than 1~ms. In the orange part of the visible spectrum, dye lasers are the only available sources which meet this specification. This is a serious drawback to the practical implementation of quantum memories using these two ions because dye lasers are relatively cumbersome systems. Moreover, their stabilization at the kHz level requires the use of an intracavity modulator to compensate for the high frequency noise induced by the fast fluctuations of the dye jet thickness \cite{Helmcke1982,Hough1984,Kallenbach1989,Julsgaard2007}.

In recent years, we have investigated several all-solid-state alternatives to dye lasers \cite{Melkonian2007,My2008-1,My2008-2,Paboeuf2011}. Among these different attempts, the intracavity frequency-doubled singly resonant optical parametric oscillator (SHG-SROPO) has given the most promising results \cite{My2008-2}. However, when dealing with  the frequency noise of optical parametric oscillators, one must be aware of a fundamental difference between these sources and usual lasers based on stimulated emission: the parametric gain process is a coherent process. Consequently, due to energy conservation at the microscopic level, the frequency fluctuations of the pump laser are transferred to the signal and idler beams, and must be shared between these two beams. In the case of the SHG-SROPO, since the pump laser is a commercial frequency-doubled Nd:YVO$_4$ laser exhibiting frequency fluctuations in the MHz range, the pump frequency fluctuations must not contaminate the useful beam, i. e., the idler beam in the present implementation. This is why we use a singly resonant architecture in which the signal beam frequency is free to fluctuate and to take away the major part of the pump frequency noise. However, the question remains of how the useful beam can be stabilized down to the kHz level. The relatively large frequency noise of the green pump that we use forbids the implementation of a stabilization of the OPO cavity on the pump frequency, which is commonly implemented by making the pump resonate in the OPO cavity \cite{Breitenbach1995,Schneider1997,Strossner1999,Petelski2001,Kovalchuk2001,Strossner2002,Kovalchuk2005,Lenhard2011}. Consequently, the frequency of the field resonant in a SROPO can be stabilized only by comparison with an optical or atomic reference and by taking effect on the cavity length. For example, different SROPOs have been stabilized at the MHz level by comparison with an interferometric \cite{Melkonian2007} or atomic \cite{Zaske2010} reference. More recently, stabilizations at the kHz level have been achieved by locking the SROPO either on the side \cite{Mhibik2010} or the peak \cite{Andrieux2011} of the transmission fringe of a Fabry-Perot \'etalon. However, in the case of lasers, it is well known that the best results are often achieved using the Pound-Drever-Hall (PDH) stabilization scheme \cite{Drever1983}. In the case of SROPOs, this scheme is interesting because it avoids the transfer of intensity to frequency noise inherent to the side of transmission fringe locking scheme \cite{Mhibik2010} and it does not require any modulation of the length of the reference \'etalon as in Ref.\ \cite{Andrieux2011}. This constitutes an important potential progress towards the use of non-tunable ultrastable reference resonators in order to improve the long-term absolute stability of the system \cite{Helmcke1987}. Consequently, the aim of the present paper is to implement a PDH locking scheme with a high finesse ($\sim 3000$) reference cavity in the case of our SHG-SROPO and to compare its performances with the side-of-fringe locking scheme implemented earlier with a much lower finesse ($\sim 100$) cavity \cite{Mhibik2010}. Furthermore, we use this opportunity to experimentally analyze how the pump noise is transferred to the non resonant wavelength.

\section{Experimental set-up}

\begin{figure}[h]
\centering\includegraphics[width=0.6\textwidth]{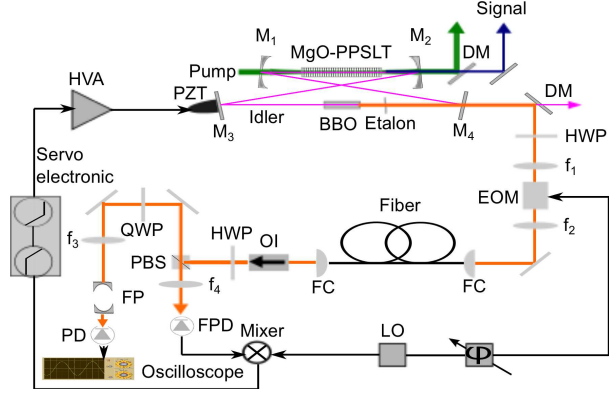}
\caption{Experimental set-up. PZT: Piezoelectric transducer; HVA: High-Voltage Amplifier; DM: Dichroic Mirror;  HWP: Half Wave Plate; LO:  Local Oscillator; f$_i$, $1\le i\le 4$: Focusing lenses; EOM: Electro Optic Modulator; FC: Fiber Coupler; OI: Optical Isolator; QWP: Quarter Wave Plate; PBS: Polarization Beamsplitter Cube;  FP: Fabry-Perot cavity;  FPD: Fast Photodiode; PD: Photodiode.}\label{fig1}
\end{figure}

Our experimental setup is schematized in Fig.\,\ref{fig1}. The OPO is pumped at 532~nm by a 10~W single-frequency Verdi laser and is based on a 30-mm long MgO-doped periodically poled stoichiometric lithium tantalate (PPSLT) crystal ($d_{\mathrm{eff}}\simeq11~\mathrm{pm/V}$) manufactured and coated by HC Photonics. This crystal contains a single grating with a period of 7.97~$\mu$m, designed to lead to quasi-phasematching conditions for an idler wavelength in the 1200-1400~nm range, leading to a signal wavelength in the 960-860~nm range. The crystal is anti-reflection coated for the pump, the signal, and the idler. The OPO cavity is a 1.15-m long ring cavity and consists in four mirrors. The two mirrors sandwiching the nonlinear crystal have a 150~mm radius of curvature. The two other mirrors are planar. This cavity is resonant for the idler wavelength only. The estimated waist of the idler beam at the middle of the PPSLT crystal is 37~$\mu$m. The pump beam is focused to a 53~$\mu$m waist inside the PPSLT crystal. All mirrors are designed to exhibit a reflectivity larger than 99.8~\% between 1.2~$\mu$m and 1.4~$\mu$m and a transmission larger than 95~\% at 532~nm and between 850~nm and 950~nm. 

In order to generate the second harmonic of the resonating idler, a 25-mm long BBO crystal is inserted between the two plane mirrors, i.~e., at the second waist of the cavity. It is anti-reflection coated for the idler and the second harmonic wavelengths. A 1.5-mm thick solid \'etalon with a finesse $F=3$ is also inserted in this leg of the cavity. The transmission of the plane mirrors is larger than 95~\% for the orange radiation. In these conditions and with the PPSLT crystal heated at $T=103\;^{\circ}\mathrm{C}$, the system oscillates at an idler wavelength of 1208~nm with a pump threshold equal to 800~mW. At a pump power equal to 3.4~W, the SHG-SROPO emits a single-frequency orange beam at 604~nm with a 30~mW output power.

\begin{figure}[h]
\centering\includegraphics[width=0.65\textwidth]{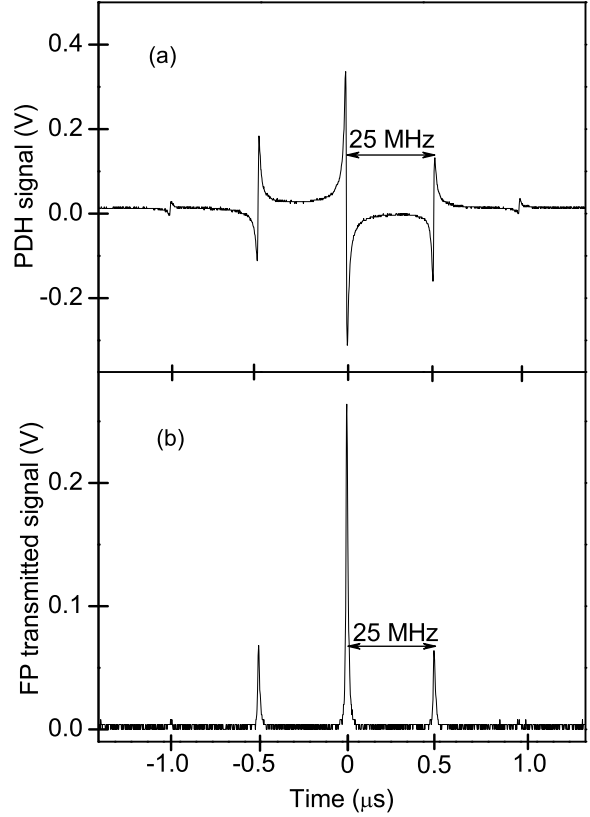}
\caption{\label{signals} (a): Pound-Drever-Hall signal. (b): Signal transmitted by the reference cavity, exhibiting the carrier frequency and the two side bands. Both signals are recorded while the length of the reference cavity is scanned.}\label{fig2}
\end{figure}

The frequency stabilization loop that we describe here is also schematized in Fig.\ \ref{fig1}. It is based on a 11.5-cm-long Fabry-Perot cavity (free spectral range equal to 1.3~GHz) with two mirrors of radii of curvature equal to 50~cm and with a finesse equal to 3000. The orange beam, which is spatially filtered and carried by a single-mode fiber, is matched to the TEM$_{00}$ Gaussian mode of this reference cavity, leading to a 78\% contrast for the reflection dip at resonance. In order to implement the PDH locking scheme, the phase of the orange beam is modulated at 25~MHz using a resonant electro-optic modulator (New Focus model 4001). The beam reflected from the reference cavity is detected on a fast photodiode (EOT model ET-2030A) and demodulated using a frequency mixer (Mini-Circuits model ZFM-3). When the phase of the demodulation is correctly chosen, one obtains the typical PDH signal \cite{Drever1983} of Fig.\ \ref{fig2}(a) when the reference cavity length is scanned thanks to a piezoelectric transducer carrying one of its mirrors. The corresponding intensity transmitted by the reference cavity is reproduced in Fig.\ \ref{fig2}(b). The two sidebands, separated by the 25~MHz modulation frequency from the carrier, are clearly seen.

\section{Experimental results and discussion}
\begin{figure}[h]
	\centering
\includegraphics[width=0.99\textwidth]{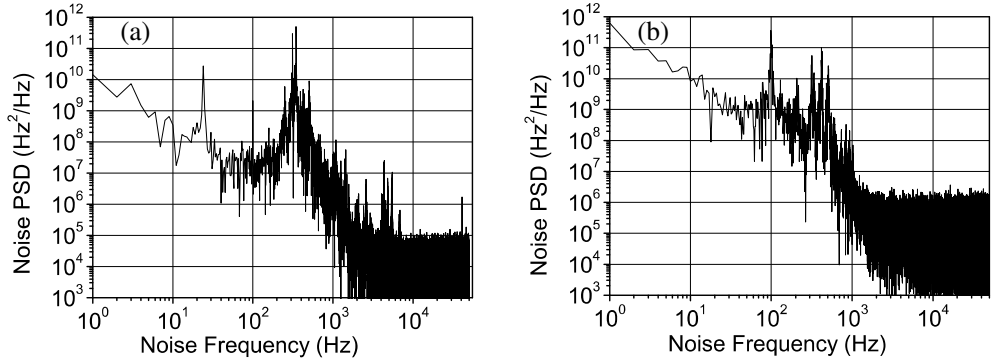}
	\caption{\label{comparaison} Spectrum of the relative frequency noise of (a) the pump laser and (b) the free-running SHG-SROPO.}\label{fig3}
	\end{figure}

Before describing the results obtained when the frequency of the second harmonic of the idler is locked, let us first discuss the free-running behavior of this frequency. Since the parametric gain is a coherent process, it is worth measuring the frequency noise spectrum of the pump laser. To perform this measurement, we send a small part of the pump beam to a confocal FP cavity with a 750~MHz free spectral range and a finesse equal to 16 at 532~nm. We manually maintain the cavity length so that the Verdi laser frequency is on the side of the transmission peak of the cavity. Then the average transmission of the cavity is one half of its transmission at resonance. In these conditions, we can record the intensity transmitted through the cavity during 1~s, which corresponds to the evolution of the frequency of the Verdi laser during 1~s. The Fourier transform of this signal gives the power spectral density (PSD) of the frequency noise \cite{Wassen1990} of the Verdi laser which is reproduced in Fig.\ \ref{fig3}(a).

The same technique is used to measure the frequency noise of the free running SHG-SROPO. A small part of the orange beam is sent to the same FP cavity, whose finesse is equal to 60 for this orange wavelength. The corresponding frequency noise spectrum is reproduced in Fig.\ \ref{fig3}(b). One can notice that the frequency noise for the orange beam is larger than for the pump, and is particularly large below 1~kHz. Notice however that for free-running lasers, when the frequency noise is large, this measurement technique cannot be considered to lead to a quantitative measurement but more to a qualitative indication of the noise bandwidth. Notice also that the noise floors at high frequency (for example above 2 kHz in Fig.\,\ref{fig3}(b)) correspond to the measurement noise and that the OPO frequency noise is actually below this floor.

To reduce this noise, we use the servo-loop schematized in Fig.\ \ref{fig1}. The PDH signal of Fig.\ \ref{fig2}(a) is filtered by a first order low-pass filter with a bandwidth of 70~kHz before being amplified using a variable gain amplifier based on an Analog Device OP-27 operational amplifier (gain bandwidth equal to 8~MHz). It is then filtered by an adjustable proportional-integrator (PI) filter (New Focus model LB1005) before being amplified by the high voltage amplifier HVA (Piezomechanik model SVR150-3) and applied to a piezoelectric transducer (Piezomechanik model PSt 150/10x10/2) carrying mirror M$_3$ of the cavity. This transducer has been specially chosen because its resonance frequencies lie above 20\ kHz. We experimentally checked using an interferometer that the combination of this HVA and this transducer with the mirror attached on it behaves like a first-order low-pass filter with a 3-dB bandwidth equal to 300\ Hz. 

\begin{figure}[h]
\centering\includegraphics[width=0.7\textwidth]{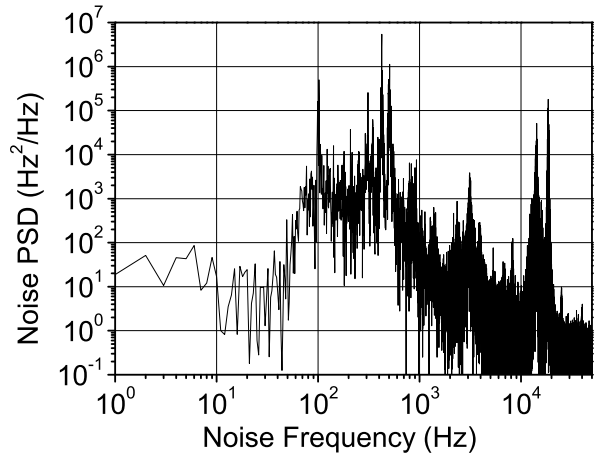}
\caption{Spectrum of the relative frequency noise of the servo-locked SHG-SROPO using the PDH technique. Notice the oscillation peaks between 10~kHz and 20~kHz.}\label{fig4}
\end{figure}

With the corner frequency of the proportional integrator filter tuned at 10 kHz, we obtain the frequency noise spectrum of Fig.~\ref{fig4} for the orange beam when the OPO is locked. In these conditions, we expect the 0~dB gain of the open-loop transfer function to lie about 5~kHz. By comparison with the free-running spectrum of Fig.~\ref{fig3}(b), we can indeed see that the low-frequency noise has been reduced by many orders of magnitude. However, we also notice the appearance of some new peaks between 10~kHz and 20~kHz. These noise peaks correspond to the onset of oscillations in our servo-loop. These oscillations come from the phase shift accumulated through the low-pass filter, the amplifier, the proportional-integral filter, the high voltage amplifier, and the piezoelectric transducer, probably combined with a residual resonance of the piezoelectric transducer, which lead to the occurrence of an overall $\pi$ phase shift at these frequencies. 

In order to get rid of this spurious oscillation, we introduced a phase-lead compensation network in the servo-loop filter. This circuit adds a $\pi/5$ phase advance at 10~kHz and induces a 15~dB gain reduction at low frequencies. It thus permits to increase the phase margin for frequencies close to 10~kHz and allows one to increase the proportional gain of the proportional integral filter by 5~dB and the corner frequency of this filter from 10 to 100~kHz while keeping the loop stable.

\begin{figure}[h]
\centering\includegraphics[width=0.7\textwidth]{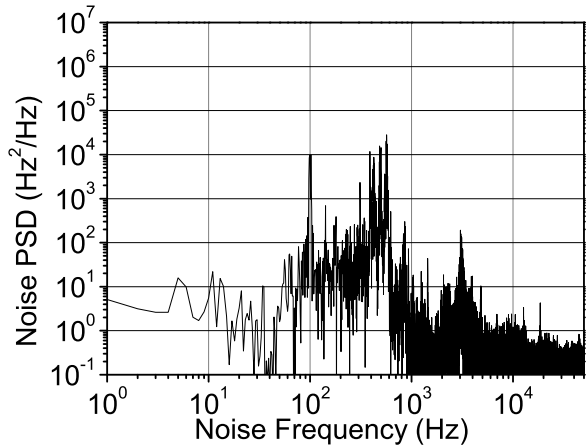}
\caption{Same as Fig.\ \ref{fig4} with a phase-lead compensator added in the loop.}
\label{fig5}
\end{figure}

In these conditions, the OPO remains locked to the reference \'etalon during several tens of minutes. One can even scan the resonance frequency of the reference \'etalon by several tens of MHz by tuning its length using a piezoelectric transducer carrying one of its mirrors while keeping the SHG-SROPO locked to this reference cavity. Analysis of the servo-loop signal leads to the relative frequency noise (i. e., frequency deviation with respect to the cavity resonance frequency) spectrum of Fig.\ \ref{fig5}. By comparison with Fig.\ \ref{fig4}, one can clearly see that the spurious resonances close to 10~kHz have disappeared. The noise is almost flat below 10~Hz$^2$/Hz, except for stronger noise components located between 100 and 1000~Hz. These noise peaks are probably due to mechanical resonances in the OPO mirror mounts, some residual oscillations of the oven containing the nonlinear crystal, and some harmonics of the 50~Hz mains frequency. When one integrates the power spectral density of the relative frequency fluctuations of the SHG-SROPO of Fig.\ \ref{fig5} from 1~Hz to 50~kHz, one obtains a rms value of 700~Hz. This corresponds to an improvement by a factor of 6 with respect to the results of ref. \cite{Mhibik2010}. It is also worth noticing that this corresponds to a rms frequency noise of 350~Hz for the idler frequency, which is the one that is actually resonating in the OPO cavity.
\begin{figure}[]
\centering\includegraphics[width=0.7\textwidth]{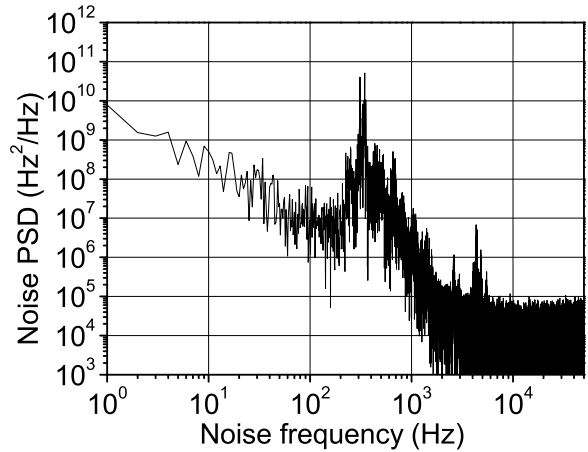}
\caption{\label{nonresonate} Spectrum of the relative frequency noise of the non-resonating signal beam while the servo-loop control of the idler frequency is on.}\label{fig6}
\end{figure}

This experiment also allowed us to clearly illustrate the transfer of frequency noise from the pump beam to the non-resonating signal beam. Indeed, comparison of the spectra of Figs.\ \ref{fig3}(a) and \ref{fig5} shows that the noise of the locked OPO is lower than the pump noise by many orders of magnitude (up to 10 at low frequencies). Since the parametric process must respect the conservation of energy $\omega_{\mathrm{p}}=\omega_{\mathrm{s}}+\omega_{\mathrm{i}}$, where $\omega_{\mathrm{p}}$, $\omega_{\mathrm{s}}$, and $\omega_{\mathrm{i}}$ are the angular frequencies of the pump, signal, and idler beam, respectively, this means that the pump frequency noise which is absent from the idler beam must be transferred to the signal beam. This is evidenced in Fig.\ \ref{fig6}, which has been obtained by analyzing the non-resonating signal beam using a confocal cavity of free spectral range equal to 1\ GHz with a finesse equal to 110. By comparison with the spectrum of Fig.\ \ref{fig3}(a), one can clearly see that the pump frequency noise, which is restrained from propagating to the idler beam by the servo-loop, is faithfully transferred to the frequency of the signal beam. Actually, the similarity of these two spectra (Fig.\ \ref{fig3}(a) and Fig.\ \ref{fig6}) is a striking illustration of the conservation of the pump photon energy in the down conversion process. Moreover, the comparison of the spectra of  Fig.\ \ref{fig3}(a) and Fig.\ \ref{fig5}, which present a huge difference, shows how the control of the idler frequency permits us to allow the transfer of the pump frequency noise only in the direction we are interested in (to the signal and not to the idler). All these features illustrate the power of the SROPO concept with respect to the DROPO concept in order to obtain an output beam with a frequency noise much lower than the pump laser and to allow the use of a relatively noisy pump laser even for applications requiring ultrahigh spectral purity.
\section{Conclusion}

In conclusion, we have been able to stabilize the frequency of an SHG-SROPO to levels lower than 1~kHz rms relative frequency noise. This has been made possible by using the Pound-Drever-Hall scheme to lock the OPO frequency to a relatively high finesse (3000) reference cavity. The pump laser noise has been experimentally shown to be almost completely transferred to the frequency of the non-resonating beam, i. e., the signal beam in the present case. This opens the way to the use of an all solid-state OPO source in rare-earth based quantum memories. Moreover, this source could also be very useful for experiments using spin coherences in nitrogen vacancy centers in diamond \cite{Hemmer2001}. Finally, such a source could be modified in order to be usable for metrology and high-resolution spectroscopy experiments in the visible, provided the reference \'etalon was made more stable on a longer time scale \cite{Helmcke1987}.

\section*{Acknowledgments}
This work was supported by the Triangle de la Physique and the Agence Nationale de la Recherche.
\end{document}